\documentclass[twocolumn,noshowpacs,preprintnumbers,amsmath,amssymb,prd]{revtex4}
\newcommand{\dt}{\ensuremath{\, \mathrm{d}}}
\newcommand{\dti}{\ensuremath{\, \mathrm{d}\kern-0.1em{}}}

\newcommand{\sign}[1]{\mathrm{sign}\left(#1\right)}

\newcommand{\abs}[1]{\left\vert#1\right\vert}

\newcommand{\expv}[1]{\langle #1 \rangle}

\newcommand{\eqrf}[1]{Eq.~(\ref{#1})}
\newcommand{\tbrf}[1]{Table~\ref{#1}}
\newcommand{\es}[1]{\, \, #1}
\newcommand{\defin}{\mathrel{\mathop:}= } 
\newcommand{\order}[1]{\ensuremath {\mathcal {O} \left( #1 \right) } }
\newcommand{\hide}[1]{\relax}

\begin{document}

\title{Quantum fields near phantom-energy `sudden' singularities}
\author{H\'{e}ctor H. Calder\'{o}n}
	\email{calderon@physics.montana.edu}
\affiliation{Department of
Physics, Montana State University, Bozeman, Montana, 59717}%
\date{\today}
\begin{abstract}
This paper is committed to calculations near a type of future singularity driven by phantom energy. At the singularities considered, the scale factor remains finite but its derivative diverges. The general behavior of barotropic phantom energy producing this singularity is calculated under the assumption that near the singularity such fluid is the dominant contributor. We use the semiclassical formula for renormalized stress tensors of conformally invariant fields in conformally flat spacetimes and analyze the softening/enhancing of the singularity due to quantum vacuum contributions. This dynamical analysis is then compared to results from thermodynamical considerations. In both cases, the vacuum states of quantized scalar and spinor fields strengthen the accelerating expansion near the singularity whereas the vacuum states of vector fields weaken it.
\end{abstract}

\maketitle
It has been determined that our universe is undergoing a second era of accelerated expansion \cite{Riess2004,Knop2003,Spergel2007,Tegmark2006,Allen2004} which can be explained by postulating the existence of a fluid whose ratio of pressure to energy density must be a negative number (see \cite{Peebles2003,Copeland2006} for reviews). This ratio is named the equation of state parameter $w \defin p/\rho$ and its effective value, after including the contributions from all fluids in the universe, must be smaller than $-1/3$ to ensure the mentioned accelerated expansion \cite{Carroll2003}. Fluids with negative $w$ are dubbed dark energy in general. The particular cases $w=-1$ and $w<-1$ are called cosmological constant and phantom energy, respectively.

Phantom energy is disfavored by theoretical reasons: it violates all energy conditions \cite{Carroll2003,Lake2004}. Yet, it is compatible with current observations albeit marginally \cite{Huterer2005}. Because the energy conditions are not satisfied, the singularity theorems do not apply and dark energy can lead to singularities. Depending on the model, phantom energy can produce future singularities that can become the end of the universe: the Big Rip singularity \cite{Turner1997,Caldwell1998}. This paper deals with softer singularities: the so-called sudden singularities \cite{Barrow2004}. This singularities are softer in the sense that divergence occurs in the second derivative of the scale factor (see \tbrf{tb:sing}). Moreover, finite objects passing through it are not crushed by tidal forces \cite{Fernandez2004}.

Some effort has gone into investigating whether the contributions from the vacuum states of quantum fields would enhance or weaken such phantom-energy singularities \cite{Calderon2005, Nojiri2005, Nojiri2004, Dabrowski2006, Barboza2006, Srivastava2007}. A word of caution is needed here. While it is possible to exploit the trace anomaly to compute the vacuum contributions (at least for conformally invariant fields in conformally flat spacetimes) \cite{Candelas1979}, it is not suitable to talk about avoiding the singularity altogether. This would imply feeding these contributions, which are of first order in $\hbar$, back to Einstein's equations. This cannot be done consistently in the models we are studying because we know only the classical behavior of dark energy, \emph{i.e.} up to zero-th order in $\hbar$.

The quantum fields considered in this paper are conformally invariant and their vacuum state is obtained by conformally transforming the Minkowski vacuum. For brevity, we will refer to them simply as ``fields'' but the mentioned adjectives are implied. On the other hand, we will use ``fluids'' to refer to classical components.

The first section of this paper is dedicated to flat FLRW with non-interacting barotropic fluids. Emphasis is given to some monotonicity results applicable if there is one dominant fluid. The second section links the behavior of the scale factor to the behavior of the equation of state parameter of a dominant barotropic fluid near a sudden singularity. It is proven that such link is necessary and sufficient. In the third section, we compute the effects of semiclassical vacuum states of conformally invariant fields. There, we show that the quantum enhancing/softening depends on the spin of the field. In the last section, general thermodynamic arguments show that the heat exchanged by these fields cannot be neglected. Thus, the models studied in this paper provide an example where the thermodynamics of gravitationally interacting systems cannot be ignored. We find that the dynamical and thermodynamical analysis coincide. Finally in the appendix, while sketching a calculation from section II, we visit some calculations of a model proposed in the literature \cite{Nojiri2005}.

\section{Flat FLRW with barotropic fluids}
The spatially flat FLRW metric (with scale factor $a$)
\begin{equation*}
  \dt s^2 =  - \dt t^2 + a^2(t)\left(\dt x^2 + \dt y^2 + \dt z^2\right)
\end{equation*}
yields the following Einstein equations:
\begin{subequations} \label{einst}
\begin{align}
  3 H^2 &= \kappa^2 \sum \rho_i\es, \label{et}
  \\
  2 \dot H &= - \kappa^2 \sum (\rho_i + p_i) \label{etd}\es,
\end{align}
\end{subequations}
where $H \defin \tfrac{\dot a} {a}$ is the Hubble rate, $\rho_i$ and $p_i$ are the energy density and the pressure of the $i$-th fluid, and the gravitational constant is set to 1 ($\kappa\defin\sqrt{8 \pi} $). If the fluids can be modeled as barotropic perfect fluids, then, by definition, their pressures are functions of their corresponding energy densities. Let us denote the negative of the enthalpy by $f$, so that
\begin{equation} \label{barot}
  f_i(\rho_i) \defin - (\rho_i + p_i)
\end{equation}
are functions of $\rho_i$ alone. In terms of $f_i$, the equation of state parameters can be written as
\begin{equation*}
  w_i =  - 1 - \frac{f_i} {\rho_i } \es.
\end{equation*}
Note that the $i$-th fluid corresponds to phantom energy as long as $f_i$ is positive definite. Using \eqrf{barot}, we can remove the pressures from Einstein's equations and from the conservation of energy equations of each fluid, which now read
\begin{subequations}
\begin{align}
  3 H &= \frac{\dot {\rho_i} + Q_i} {f_i} \label{cem}\es,\\
  \sum Q_i &= 0\es.
\end{align}
\end{subequations}
The $Q_i$ account for the transfer of energy between fluids. As usual, conservation of energy and Einstein's equations are not independent.

Now, we will make the assumption that there is no interaction amongst the fluids, i.e. $Q_i = 0$. This step will be justified by its compatibility with the results: If at the end of the calculations there is only one dominant fluid, phantom energy near the singularity, whose energy density is much larger than the energy densities of all other fluids combined, then we can neglect the interaction between fluids. In this case, we can use \eqrf{cem} to find $\rho_i$ as functions of $a$, and \eqrf{et} to find the evolution of $a$ in time:
\begin{align}
  3\log\left( \frac{a} {a_0}\right) &= \int^{\rho_i}_{ {\rho_i}_0 } \frac{\dt
  \rho_i} {f_i(\rho_i)} \label{rfa}\es,
  \\
  \frac{\kappa} {\sqrt{3}}(t - t_0) &= \int^a_{a_0} \frac{\dt a} {a\sqrt{\sum
  \rho_i(a)}}\es. \label{tfa}
\end{align}
The subindex zero indicates the value of the quantity at a given time, say today.

From \eqrf{rfa}, we can reason that $\rho_i$ are monotonic functions of $a$ as long as $f_i$ don't cross zero, cross infinity or jump branches ($f_i$ could be multivalued). Analyzing \eqrf{tfa}, we can deduct that $a$ is a monotonic function of $t$ provided that the total energy density $\sum \rho_i$ does not vanish or diverge. We can conclude that $\rho_i$ are monotonic functions of time until $f_i$ or $\sum \rho_i$ vanish or diverge or $f_i$ change branches.

Finite-time future singularities will appear if for some reason the integral in \eqrf{tfa} cannot keep increasing; so that time, in the left hand side of equation of \eqrf{tfa}, wouldn't keep increasing either. A classification of finite-time future singularities was introduced by Nojiri \emph{et al.} \cite{Nojiri2005} and completed by Copeland \emph{et al.} \cite{Copeland2006}. It is summarized in \tbrf{tb:sing}.

If the integrand in \eqrf{tfa} cannot be evaluated, say because the integrand fails to be a real number for $a = a_s$, then we might have a sudden or type-III singularity at $a_s$. On the other hand, if $a$, the upper limit of \eqrf{tfa}, can go to infinity, then we have a Big Rip singularity if the integral converges; and there is no singularity if the integral diverges.

One corollary out of these statements is that a singularity type IV cannot be produced by a single barotropic non-interacting fluid in a realistic model. This case, although mathematically possible, is physically impossible. Baryonic matter will always be present and its contribution to the right hand side of \eqrf{et}, which evolves as $a^{-3}$, does not vanish because $a_s$ remains finite per definition of type-IV singularity. If the energy density of phantom fluid vanishes then the energy density of baryonic matter takes over and drives the evolution of $a$ (without any singularity whatsoever).

\begin{table}
  \caption{\label{tb:sing} Classification of finite-time future singularities \cite{Copeland2006,Nojiri2005}. A dash indicates non-specified behavior. The equivalences of behavior, $\rho \sim\abs{\dot{a}}$ and $\abs{p}\sim\abs{\ddot{a}}$, follow from Eqs.~(\ref{einst}). }
  \begin{ruledtabular}
	\begin{tabular}{l|cccc}
	  & $a$ & $\rho\sim\abs{\dot{a}}$ & $\abs{p}\sim\abs{\ddot{a}}$ & $\abs{\dddot{a}}$ and higher\\
	  \hline
	  I - Big Rip & $\infty$ & $\infty$ & $\infty$ & - \\
	  III         & $a_s$ & $\infty$ & $\infty$ & - \\
	  II - sudden & $a_s$ & $\rho_s$ & $\infty$ & - \\
	  \hline
	  IV          & $a_s$ &  $0$  & $0$ &  $\infty$  \\
	  V           & $\infty$ &  $\rho_s$  & $p_s$ &  $\infty$  \\
	\end{tabular}
  \end{ruledtabular}
\end{table}

\section{Sudden singularities}
A sudden singularity occurs when the pressure diverges but the energy density and the scale factor remain finite. By \eqrf{einst}, the first derivative of the scale factor also remain finite but the second derivative diverges. Hence, $a$ can be approximated, near the singularity, by
\begin{equation} \label{aaprox}
  a(t) \approx a_s - A \, (t_s - t) + B \, (t_s - t)^{1 +  \frac{1} {1 + \delta}} +
  \mathcal{O}((t_s - t)^{1 +  \frac{1} {1 + \delta} + \epsilon} )
\end{equation}
with positive definite both $\delta$ and $\epsilon$. Possible higher order terms in $t_s - t$ have been omitted ($\epsilon > 0$) because we only need to show the divergence of the second derivative while keeping the first derivative finite ($\delta > 0$). It is easy to prove that the behaviors of the total energy density and total pressure near the singularity are given by
\begin{subequations}
\begin{align}
  \sum \rho_i &\approx \frac{3\, A^2}{a_s^{\phantom{s}2}\, \kappa ^2}  - \frac{6
  \,A\, B\, (\delta +2) \,(t_s - t)^{\frac{1}{1 + \delta}}} {a_s^{\phantom{s}2}\,
  (\delta + 1)\, \kappa ^2}\es,\label{eq:EneBeh}\text{ and}
  \\
  \sum p_i &\approx - \frac{2 \,B\, (\delta +2)}{a_s (\delta + 1)^2} \,(t_s - t)  ^{
  - 1 + \frac{1}{1 + \delta}}\es.
\end{align}
\end{subequations}
Note that the signs in front of $A$ and $B$ in \eqrf{aaprox} have been chosen in such a way that if $A$ and $B$ are both positive then $a$ approaches $a_s$ from below, $\rho$ also approaches $\rho_s$ from below, and $p$ diverges to $-\infty$. If $A$ was negative, then $a'(t_s)$ would be negative. But this cannot happen because of the monotonicity of $a$  --Einstein equations for flat FLRW don't have curvature terms to change the sign of $a'(t)$-- and we know that $a'(t)$ is positive today.

Now, let us analyze the conditions under which the contributions from fluids like dark matter or electromagnetic radiation would not be significant near the singularity. We need the total energy density to be much larger than the dark-matter energy density. This is, the constant term in the behavior of $\sum \rho_i$, \emph{cf.} \eqrf{eq:EneBeh}, must be much larger than ${\rho_m}_0 a_0^{\phantom{0}3}/a_s^{\phantom{s}3}$, where ${\rho_m}_0$ is the energy density of dark matter measured today. Therefore, $a_s$ and $A$ must be big enough to satisfy
\begin{equation*}
  A^2a_s \gg  \frac{\kappa^2} {3}\,a_0^{\phantom{0}3}\,{\rho_m}_0\,.
\end{equation*}
If $A$ vanished, then this inequality could not be satisfied. This is, only positive definite $A$ is compatible with the assumption of non-interacting fluids.

If $B$ was negative, the fluid causing the singularity would have positive both energy density and pressure. Thus, such fluid would, near the singularity, simply not be phantom energy.

Assuming that only one fluid, dark energy, contributes significantly near the singularity, then \eqrf{barot} for such fluid has the form
\begin{equation} \label{fappAB}
  f \approx \sign{B}\frac{(3 \,\abs{A})^\delta(2\, \abs {B} \, (\delta +2))^{1 + \delta}
   }{\kappa ^ {2 \delta}\,(\delta +1)^{2 + \delta} \,a_s^{\phantom{s}1 + 2 \delta}}
   \,\abs{\frac{3 A^2}{a_s^{\phantom{s}2} \kappa ^2} - \rho}^{ - \delta}.
\end{equation}
Although it has been argued that only positive $A$ and $B$ are physically relevant for phantom-energy driven future singularities, the above formula shows the correct sign dependence should $A$ or $B$ be negative. Keeping the sign dependence allows for easier comparison with other calculations presented in the literature.

Conversely, if phantom energy is modeled by
\begin{equation} \label{fsud}
  f = \frac{C} { (\rho_s - \rho) ^\delta} + \mathcal{O}((\rho_s - \rho)^{1 -
  \delta})\,,
\end{equation}
with positive $C$, $\rho_s$, and $\delta$, then the evolution is such that the scale
factor near the singularity is given by
\begin{widetext}
\begin{equation}
  a(t) \approx a_s \left( 1 - \tau + \frac { \left( \frac {3 \,C \, ( 1 + \delta ) } { \kappa^2} \right)^{ \frac {1} { 1 + \delta } } ( 1 + \delta ) } {2 \,( 2 + \delta ) \, \rho_s} \, \tau^{1 + \frac {1} {1 + \delta} } + \order { \tau^{1 + \frac {1} {1 + \delta } + \epsilon } } \right) \label {atrunc}
\end{equation}
\end{widetext}
where
\begin{equation}
   \tau \defin \kappa \sqrt {\frac {\rho_s} {3} } ( t_s - t ) \es, \label {deft}
\end{equation}
and
\begin{equation}
  \epsilon = \min\left\{2, \frac{\delta + 3} {\delta + 1}\right\} - 1 - \frac{1} {1 +
  \delta}\es.
\end{equation}
Note that \eqrf{atrunc} is of the form of \eqrf{aaprox} and that $\epsilon$ is positive for $\delta > 0$. While Copeland \emph{et al.} showed that the first term of \eqrf{fsud} yields a sudden singularity (see Eq.~(461) in \cite{Copeland2006}), the calculation shown here is more general in that it only analyzes the behavior near the singularity (hence the operator $\mathcal{O}$ and the need to keep track of $\epsilon$). Thus, it encompasses other models, {\it e.g.} model (32) in \cite{Nojiri2005}, that might behave differently far from the singularity. As implied in remarks by Catt{\"o}en and Visser \cite{Cattoen2005}, proving that \eqrf{fappAB} can be obtained from \eqrf{aaprox} is relatively trivial. However, the proof of the converse is rather laborious because one must prove that the dismissed terms in \eqrf{fsud} can also be dismissed in \eqrf{atrunc}. Such proof follows the lines of the calculation in the appendix.

We reach then the following conclusion: a phantom-energy model where barotropic dark energy is the only significant fluid near the singularity will produce a sudden singularity, \eqrf{aaprox}, if and only if its behavior near the singularity has the form of \eqrf{fsud}. The relationship between $(A, B)$ and $ (C, \rho_s) $ can be read off from equations (\ref{aaprox}) and (\ref{atrunc}). One implication of this theorem is that sudden singularities cannot be achieved with a static equation of state parameter, it must evolve according to
\begin{equation} \label{wend}
  w \approx - \frac{C} { \rho\, (\rho_s - \rho) ^\delta} = \mathcal{O}(\tau ^{ - 1 +  \frac{1} {1 + \delta }})
\end{equation}
near the singularity.

\section{Semiclassical fields}
\begin{table}
  \caption{\label{tb:cof} Spin-dependent coefficients in \eqrf{tmn}.}
  \begin{ruledtabular}
	\begin{tabular}{ccc}
	  spin & $\alpha$ & $\beta$
	  \\
	  \hline
	  0 & $\frac{1} {2800\pi^2}$ & $\frac{1} {2800\pi^2}$
	  \\
	  $ \frac{1} {2}$  & $\frac{3} {2800\pi^2}$& $\frac{11} {5600\pi^2}$
	  \\
	  1 & $ - \frac{9} {1400\pi^2}$ & $\frac{31} {1400\pi^2}$
	\end{tabular}
  \end{ruledtabular}
\end{table}
We will use the semiclassical expression for the vacuum stress-energy of conformally invariant quantized fields in a vacuum state conformally obtained from Minkowski spacetime \cite{Candelas1979}:
\begin{multline} \label{tmn}
  \expv{T_{\mu\nu}} =  \frac{\alpha} {3}\left( g_{\mu\nu} \,
  R^{;\sigma}_{\phantom{;\sigma};\sigma} - R_{;\mu\nu} + R\,R_{\mu\nu} -  \frac{1}
  {4}g_{\mu\nu}\,R^2\right) +\\ \beta\left( \frac{2} {3}R\, R_{\mu\nu} -
  R^{\sigma}_\mu\,R_{\nu\sigma} +  \frac{1} {2} g_{\mu\nu}\, R_{\sigma \tau }\,R^{\sigma \tau } -  \frac{1} {4}g_{\mu \nu }\,R^2 \right)\,,
\end{multline}
where $R$ is the Ricci scalar, $R_{\mu\nu}$ is the Ricci tensor, and $\alpha$ and $\beta$ are spin-dependent coefficients given in \tbrf{tb:cof}. Because the derivatives of $R$ diverge faster than $R$ or $g$, the first two terms of the coefficient of $\alpha$ are the ones that contribute the most:
\begin{widetext}
  \begin{subequations} \label {eq:rhopSudden}
  \begin{align}
    \rho_a \defin  \expv{T_{00}}  \vert_{\text {spin} = a}  & = \alpha \frac { \kappa^4 \, \rho_s} {3} ( 3 \,C )^{ \frac {1} {1 + \delta }} \, \delta \left( ( 1 + \delta ) \, \tau \right)^{- 2 + \frac {1} {1 + \delta}} \es ,
    \\
    p_a \defin \expv{T_{11}}  \vert_{\text {spin} = a} & = - \alpha \frac {\kappa^4 \, \rho_s} {9} ( 3 \,C )^{\frac {1} {1 + \delta} } \, \delta \,( 1 + 2 \,\delta )\,  {a_s}^2 \,\left( ( 1 + \delta ) \, \tau \right)^{- 3 + \frac {1} {1 + \delta}} \es . \label{eq:exppSudden}
  \end{align}
  \end{subequations}
\end{widetext}
Both expectation values diverge as the singularity approaches, the pressure faster than the energy density. The equation of state parameter of the quantum contributions is then given by
\begin{equation}
  w = - {a_s}^2 \frac {1 + 2 \, \delta} {3 \,( 1 + \delta ) } \tau^{ -1} \es .
\end{equation}
It is negative, it doesn't depend on the spin of the field, and it diverges at the singularity faster than \eqrf{wend}. These divergences mean that the approximation breaks down before the singularity occurs. Nevertheless, a qualitative analysis is in order. The behavior of the perturbation is determined by the sign of $\alpha $, which is the only quantity not defined positive in \eqrf{eq:exppSudden}. Because $\alpha$ is positive for both scalar and spinor fields, then $\rho_a >0$, $p_a<0$ and therefore these vacuum states enhance the singularity. This happens because adding these vacuum perturbations is equivalent to adding more phantom energy to the right hand side of \eqrf{einst}. Vector fields, with negative $\alpha$, counter the contributions from dark energy and therefore soften the singularity.

\section{Thermodynamical Considerations}
While the vacuum state of a system described by an exact Lagrangian has no thermal properties, it is not impossible for a vacuum state to become thermal (\emph{e.g.} \cite{Mitter2000, Horwitz1986}). In general (see \cite{Katz1967} for a canonical exposition), small unknown terms in the Hamiltonian cause the thermalization of the system (equation (12.2) leads to equation (13.13) in the reference).

In this section, the vacua are considered as individual subsystems. We admit that their evolution is known only up to some terms of first order in $\hbar$ and that their complete Lagrangian would contain higher order terms in $\hbar$, self-interactions, and interactions with the phantom fluid. Thus, these vacua qualify for thermal calculations. The previous section provide the energy density and pressure of the fields. The pressure is interpreted as the partial pressure of the vacuum state in the mixture of cosmological fluids. We don't need to assume that the process is quasi-static because we are not concerned with the temperature or the entropy of the subsystems; we shall be satisfied with computing the sign of the change of the enthalpy of formation.

In the first law of thermodynamics $\delta U = \delta Q - \delta W$, we can replace
\begin{subequations}
\begin{align}
  U & \propto \rho_a \, a^3\,,\\
  W & \propto p_a \, a^3 \,. \label{wprop}
\end{align}
\end{subequations}
The approximations for \eqrf{tmn} are valid only for free fields (see \cite{Birrell1984} page 5). Thus, the lack of interaction terms in \eqrf{wprop} is justified. We are interested in the sign of $\delta Q$ because it determines whether the expansion is exothermic or endothermic. As shown in Eqs.~(\ref{eq:rhopSudden}), the pressure diverges faster than the energy density and therefore $\delta Q \approx \delta W$. Hence,
\begin{equation}
  \sign {\delta Q} = \sign {3\,  p_a\, a^2 \, \delta a + a^3 \, \delta p_a}\,.
\end{equation}
We can now dismiss the first term in the right hand side because it behaves as $\tau^{-3 + \frac {1} {1 + \delta} }$ which is slower than the $\tau^{-4 + \frac {1} {1 + \delta} }$ divergence of $\delta p_a$. Therefore, the sign of the heat flowing \emph{into} the system is the same as the sign of pressure change. The latter is determined by the negative of the sign of $\alpha$ because all of the other quantities in \eqrf{eq:exppSudden} are positive definite. The expansion is then exothermic for both scalar and spinor fields and endothermic for vector fields. Exothermic reactions are spontaneous and thus they enhance the singularity. We conclude again that scalar and spinor vacua enhance the singularity and vector vacua soften it. This is in agreement with the dynamical results of the previous section.

While the models considered in this paper furnish an example of dynamics and thermodynamics coinciding in their predictions, this might not necessarily be true in general. A counterexample would be extremely valuable because it will point to a deficiency of Quantum Field Theory in Curved Spacetimes (QFTCS). This theory is considered complete up to first order in $\hbar$. But if the dynamics and thermodynamics don't coincide, then the predictive power of QFTCS would be limited because it wouldn't be possible to determine the evolution of an entropy-driven system without a Statistical Mechanics theory of gravity (let alone Quantum Gravity).

\appendix*
\section{Analysis of model (32) of \cite{Nojiri2005}}
In this appendix we will follow the steps necessary to obtain \eqrf{aaprox} from \eqrf{fsud}. The general case is rather complicated because of the need to keep correct track of the orders of magnitude, and at some point it involves expanding a hypergeometric function composed with a logarithm evaluated at the singularity of the logarithm. So, instead of cluttering with long mathematical expressions, we will perform the basic steps while reviewing calculations in a model proposed by Nojiri \emph{et al.} (model (32) of \cite{Nojiri2005}). Also, this derivation will expose some problematic points in case the reader is interested in reproducing the full calculations.

Consider dark energy modeled as a fluid with
\begin{equation}
  f(\rho) = \frac{b\,\rho^{1 - \gamma}} {\gamma\left(\rho_s^\gamma -
  \rho^\gamma\right)} \label{deff}
\end{equation}
where $\rho_s > 0$ is the dark energy density at the singularity \footnote{Parameters $A$, $B$, $\alpha$, and $\beta$, in equations (32) and (35) of \cite{Nojiri2005}, are related to $\rho_s$, $b$, and $\gamma$ by $A = -b / \gamma$, $B = b\,\rho_s^{ - \gamma} / \gamma$, and $\beta = 1 - \gamma$.}, and $\gamma \neq 0$. If $\rho_s$ was negative, then $f$ might not be defined over the real numbers. Using the monotonicity results from section I, it is trivial to show that \eqrf{deff} corresponds to phantom energy as long as and whenever $\rho_0 < \rho_s$. On the other hand, if $\rho_0 > \rho_s$, the fluid described by \eqrf{deff} might begin its evolution as non-phantom dark energy (for $\gamma > 0$ and $\rho_0 > \rho_s\big(1 + \sqrt{1 + 4\,b/\gamma \rho_s^{\phantom{s}2\,\gamma}}\,\big)/2$) or a normal fluid (i.e. positive both pressure and energy density) but it will become a normal fluid near the singularity.

With this model, \eqrf{rfa} yields
\begin{equation}
  \rho^{\gamma} = \rho_s^{\phantom{s}\gamma} - \left(  \rho_s^{\phantom{s}\gamma} -
  \rho_0^{\phantom{0}\gamma} \right) \sqrt{1 - b \frac{\log\left( \frac{a}
  {a_0}\right)} {\left(\rho_s^{\phantom{0}\gamma } - \rho_0^{\phantom{s}\gamma}
  \right)^2}}
\end{equation}
Even though the above equation might have two signs at front of the radical sign, one of them is spurious. The sign shown yields $\rho\rightarrow \rho_0$ when $a\rightarrow a_0$.

As argued previously, the singularity occurs when the quantity inside the square root vanishes. This defines the value of $a$ at the singularity:
\begin{equation}
  a_s = a_0 \exp\left( \frac{\left(\rho_s^{\phantom{s}\gamma} -
  \rho_0^{\phantom{0}\gamma}\right)^2} {6\,b}\right) \es .\label{as}
\end{equation}
If $b$ is negative, then the radius at the singularity will be smaller than the radius today. In that case, the contribution from matter and dark matter is not diluted and a model without such contribution is unphysical. We will continue our analysis assuming $b>0$.

Since we are interested in the behavior near the singularity, instead of \eqrf{tfa},
\eqrf{et} can be approximately written as
\begin{equation}
  \frac{\kappa} {\sqrt{3}}(t_s - t) \approx \int^{a_s}_a \frac{\dt a} {a\sqrt{\rho(a)}}
\end{equation}
The contributions from other fluids but dark energy have been discarded on the basis that $a_s$ is much larger than $a_0$. The integrand can be expanded around $a_s$, the integral can be evaluated, and the resulting series can be inverted. The first three terms of the outcome are
\begin{equation}
  a(t) = a_s\left(1 - \tau - \lambda \sqrt{ \frac{2\,b} {3}}\,
  \frac{\rho_s^{\phantom{s} - \gamma} } {\abs\gamma}\,\tau^{ \frac{3} {2}}\right) +
  \mathcal{O}\left(\tau^2\right)
\end{equation}
where $\lambda \defin \sign{\log(\rho_0/\rho_s)}$ and $\tau $ as in \eqrf{deft}.

Note that the second derivative of $a$ is the first to diverge. Thus, the singularity is only a sudden singularity, no matter the value of $\gamma$. \hide{The rich structure reported in \cite{Nojiri2005} is derived from both mathematical errors (e.g. not tracking the constants appropriately) and errors in the physical model (e.g. not accounting for the contribution of other fluids when $a_s$ is not much larger than $a_0$).} Contrast this to the rich $\gamma$-dependent structure reported elsewhere \cite{Nojiri2005}.

The independence from $\gamma$ could also be deduced from the behavior of \eqrf{deff} near the singularity:
\begin{equation}
  f(\rho)\approx  \frac{b\,\rho_s^{\phantom{s}2(1 - \gamma)}} {\gamma^2 \abs{\rho_s - \rho}} + \mathcal{O} \left( (\rho_s - \rho)^0 \right)
\end{equation}
which shows that $\rho_s$ is a single pole and therefore falls within the theorem of section II.


\end{document}